\documentclass[conference]{IEEEtran} 
\usepackage{amsmath}
\usepackage{amsmath,amssymb,amsfonts,amsthm}
\usepackage{cite}
\usepackage{mathtools}
\usepackage{algorithm}  
\usepackage{bm}   
\usepackage[binary-units=true]{siunitx}
\DeclareSIUnit{\belmilliwatt}{Bm}

\usepackage{graphicx}
\usepackage{dsfont}
\usepackage{subcaption}
\usepackage{comment}

\usepackage{textcomp}
\usepackage[export]{adjustbox}
\usepackage{float}
\usepackage{booktabs}
\usepackage{multicol}
\usepackage{xparse}
\usepackage{color,soul}
\usepackage[bottom]{footmisc}
\usepackage[binary-units=true]{siunitx}
\DeclareSIUnit{\belmilliwatt}{Bm}
\DeclareSIUnit{\dBm}{\deci\belmilliwatt}
\DeclareSIUnit[per-mode=symbol,per-symbol=p]{\Bps}{\byte\per\second}

\sethlcolor{yellow}
\usepackage{algpseudocode}

\usepackage{hyperref}


\IEEEoverridecommandlockouts
\begin{document}

    \title{Intelligent Radio Resource Slicing for 6G In-Body Subnetworks}

\author{
\IEEEauthorblockN{Samira Abdelrahman\IEEEauthorrefmark{1} and Hossam Farag\IEEEauthorrefmark{2}\IEEEauthorrefmark{1}}
\IEEEauthorblockA{
\IEEEauthorrefmark{1} Department of Electrical Engineering, Aswan University, Egypt\\
\IEEEauthorrefmark{2}Department of Electronic Systems, Aalborg University, Denmark \\
Email: sma@asw.edu.eg,  hmf@es.aau.dk
}
}
	\maketitle
	\begin{abstract}
6G In-body Subnetworks (IBSs) represent a key enabler for supporting standalone eXtended Reality (XR) applications. IBSs are expected to operate as an underlay to existing cellular networks, giving rise to coexistence challenges when sharing radio resources with other cellular users, such as enhanced Mobile
Broadband (eMBB) users. Such resource allocation problem is highly dynamic and inherently non-convex due to heterogeneous service demands and fluctuating channel conditions. In this paper, we propose an intelligent 
radio resource slicing strategy based on the Soft Actor-Critic (SAC) deep reinforcement learning algorithm. The proposed SAC-based slicing method addresses the coexistence challenge between IBSs and eMBB users by optimizing a refined reward function that explicitly incorporates XR cross-modal delay alignment to ensure immersive experience while preserving eMBB service guarantees. Extensive system-level simulations are performed under realistic network conditions and the results demonstrate that the proposed method can enhance user experience by 12–85\% under different network densities compared to baseline methods while maintaining the target data rate for eMBB users.
	\end{abstract}
\section{Introduction}\label{sec:intro}
The sixth generation (6G) of wireless networks is envisioned as a network-of-networks~\cite{6G}, where heterogeneous radio access technologies, deployment scales, and service paradigms coexist and are jointly orchestrated to support extreme performance requirements. A key pillar of this vision is the concept of 6G in-X subnetworks~\cite{in-x}, which provide short-range, low-power, and high-performance wireless connectivity within physical entities such as industrial modules, vehicles, robots, and the human body. By operating at very short distances and leveraging localized coordination, in-X subnetworks are expected to support services with stringent latency, reliability, and availability requirements that have traditionally relied on wired connectivity. Using wireless for such applications avoids the drawbacks related to a wired setup, including higher cost, limited deployment flexibility, and maintenance of cables. In-Body Subnetworks (IBS)~\cite{in-x, ibs} play a pivotal role in supporting proximity wireless communications around the human body, with eXtended Reality (XR) standing out as a principal application scenario. 

From a practical deployment perspective, in-X subnetworks typically operate under the coverage of a larger network~\cite{in-x}, such as a 5G or beyond cellular system, where they act as underlay networks and share radio resources with conventional cellular users. For instance, IBSs may coexist with conventional cellular users, reusing time–frequency–spatial resources managed by the  radio access network (RAN). XR service requirements are located between ultra-reliable low-latency communications (URLLC) and enhanced mobile broadband (eMBB)~\cite{traffic}. Hence, such coexistence introduces radio resource management (RRM) challenges that extend beyond traditional heterogeneous networks. The large disparity in transmit power and coverage between cellular base stations and in-X access points results in highly asymmetric interference, where cellular transmissions can jeopardize the reliability of in-X links, while aggregated interference from multiple subnetworks may degrade eMBB (cellular users) performance. Moreover, coexistence must accommodate heterogeneous service requirements of both in-X subnetworks alongside high data rates of eMBB cellular users. The interference environment is highly dynamic, driven by entity-specific subnetwork activity and time-varying cellular traffic. A fundamental requirement pertains to IBS-empowered XR application is the synchronized video-haptic transmission to ensure immersive user experience. This demands a resource allocation mechanism capable of accommodating the distinct latency and rate constraints of each modality while preserving inter-modal synchronization.

Motivated by the aforementioned challenges, this work aims to introduce an intelligent radio slicing Soft Actor-Critic (SAC) deep reinforcement learning (DRL) algorithm. We address the scenario where the IBSs underlay a macro cellular network and share the same radio resources with eMBB users. The proposed method adopts a refined reward function that incentivizes the slicing mechanism to fulfill Quality-of-Service (QoS) objectives of each slice and dynamically adapts to changes in the network conditions. We first consider the inter-slice distribution where the available radio resource are distributed among IBS and eMBB users considering data rate, packet loss and synchronization requirement. Then, we consider intra-slice scheduling where the resources allocated for each user set (slice) are fairly distributed considering the buffer occupancy. The effectiveness of the proposed method is evaluated via system-level simulations and the results demonstrate significant performance gains in terms of user satisfaction ratio (12–85\%) compared to relevant baseline while maintaining the target throughput of eMBB users.

The paper is organized as follows. Section~\ref{sec:related} describes the related work. Section~\ref{system-model} presents the proposed DRL-based radio slicing method. Performance evaluations are given in Section~\ref{results} and finally the paper is concluded in  Section~\ref{sec:conclusions}.
\section{Related Work}\label{sec:related}
Existing works on in-X subnetworks have mainly focused on addressing the challenges of radio resource management and interference mitigation. In~\cite{RNN},  a multi-agent DRL framework for resource scheduling in non-coordinated in-X subnetworks, combining recurrent neural network and a binary tree search procedure to cope with time-varying channel conditions. The work in~\cite{goal} addresses decentralized interference management in industrial subnetwork deployments by introducing a goal- and control-aware coordination algorithm tailored for subnetwork-controlled plants. The authors in~\cite{proactive} introduces a proactive radio resource allocation method using Bayesian ridge regression to minimize the age of information in industrial subnetworks. However, these works have assumed dedicated spectrum availability for in-X subnetworks and have not addressed their coexistence with cellular networks where sub-networks share resources managed by a cellular entity. While coexistence has been explored in similar comparable settings, such as device-to-device settings~\cite{d2d1, d2d2}, these studies do not explicitly account for the unique coexistence challenges arising from the underlay deployment of in-X subnetworks.

RAN slicing~\cite{RAN-slicing} naturally emerges as a key abstraction for managing coexistence, enabling logical isolation between cellular users and in-X services while allowing controlled resource sharing. In the underlay coexistence scenario, slicing decisions must be continuously adapted to the evolving interference landscape and heterogeneous service demands. In addition, the capability of dynamically adapting resource slicing is therefore of paramount importance or achieving the expected communication requirements of in-X subnetworks under high mobility and dynamic subnetwork crowds~\cite{crowds}. Therefore, simple heuristic-based methods such as equal resource splitting across slices are inefficient, as they ignore slice heterogeneity and complex network dynamics~\cite{Heur}. Moreover, the problem of determining the optimal resource allocation at each system state is NP-hard, thereby limiting the applicability of theoretical optimization methods~\cite{NP}. This necessitates the need for intelligent, adaptive, and learning-based RRM frameworks capable of dynamically configuring slices and allocating radio resources in a way that balances isolation, efficiency, and robustness. Several data-driven methods have been investigated in the literature~\cite{ran1, ran2, base}. However, these works focus mainly on maximization/minimization of specific network metrics for each slice and do not consider the corresponding QoS objectives/constraints for each slice, leading to either degraded user experience or resource over-provisioning. Moreover, none of these works consider the cross-modal synchronization for XR uses cases supported by IBSs.
\section{The proposed DRL-Based RAN Slicing}
\label{system-model}
    \begin{figure}[t!] 
		\centering
		\includegraphics[width= 1\linewidth]{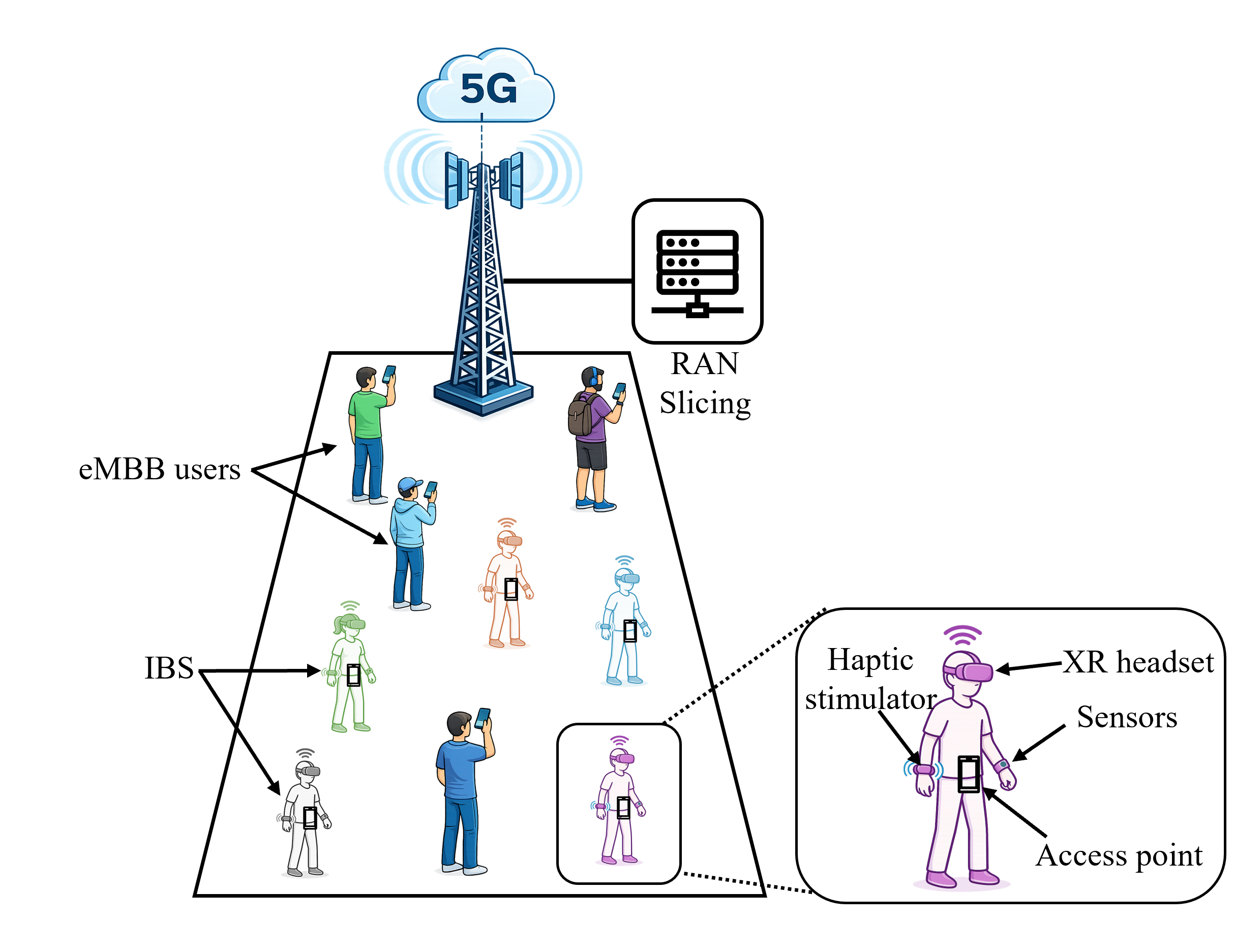} 
		\caption{Network model showing a number of IBSs (XR users) coexist and share the radio resources with eMBB users.\label{network}}
	\end{figure}
We consider the scenario of a set $\mathcal{N}=\{1, 2, ..., N\}$ of consumer subnetworks underlay a 5G macro base station and share the same resources with a set set $\mathcal{M}=\{1, 2, ..., M\}$ of eMBB cellular users as depicted by Fig.~\ref{network}. The network operates using orthogonal frequency-division multiple access (OFDMA) with subcarrier spacing determined by the numerology of 5G-NR. The total network bandwidth is divided into a number resource blocks (RBs) and these RBs are grouped into $R$ RB groups (RBGs). The macro base station is responsible for resource management and RAN slicing for both the eMBB and IBS users. The traffic pattern as well as the QoS requirements are the same for all users within the same set. The eMBB users use a full-buffer traffic pattern, requiring resources for transmission at each time slot. For the IBS users, we adopt single-eye-buffer traffic model~\cite{traffic}, where video frames for both eyes arrive together as one application-layer packet.  We focus on the downlink transmission, where XR devices receive high-bandwidth video frames rendered at the subnetwork access point (AP). Specifically, the AP collects data from sensors and controllers, computes appropriate interactive video frames, and transmits them back to the XR headset. Another option could be that the XR headset has high capability of computation so that the XR scene can be self-generated~\cite{synch}. For immersive experience, the XR users are provided with a haptic feedback in the form of vibration or heat. Such feedback shall be tightly synchronized with the XR scene delivered to the users. In our work, we focus on three network requirements: data rate, packet loss, latency and video-haptic synchronization. 

\begin{figure}[t!] 
		\centering
		\includegraphics[width= 1\linewidth]{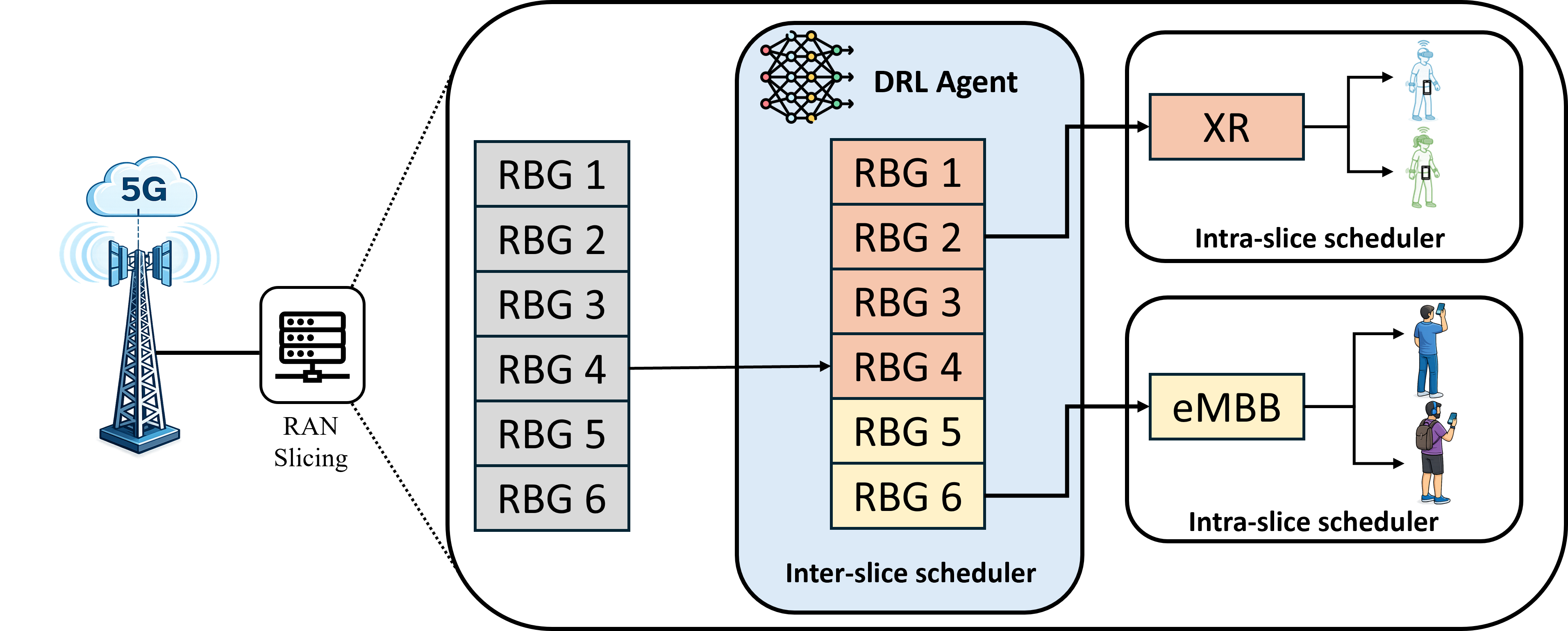} 
		\caption{The DRL-based RAN slicing showing inter-slice and intra-slice schedulers.\label{slicing}}
	\end{figure}

The aim of the proposed DRL-based RAN slicing algorithm is to allocate RBGs to a slice $k\in \{s,c\}$, where $s$ denotes the XR slice and $c$ denotes the eMBB-slice as depicted by Fig.~\ref{slicing}. For the considered complex scenario of a heterogeneous, time-varying network, we employ DRL-based policy that dynamically allocates RBGs for each slice. Particularly, we adopt the Soft Actor–Critic (SAC) algorithm due to its strong performance in highly dynamic environments. SAC maximizes a stochastic policy objective that jointly considers the expected long-term return and an entropy regularization term, where the entropy coefficient regulates the balance between exploration and exploitation. During training, the agent stores transition tuples, comprising the current state, selected action, obtained reward, and subsequent state in a replay buffer, from which mini-batches are randomly sampled to update the actor and critic networks~\cite{SAC}. As SAC inherently produces continuous-valued actions, the generated outputs are discretized to determine the number of RBGs allocated to each slice $k$, following a similar approach deployed in~\cite{slicing}. The agent, state, action and reward of the DRL method are defined as follows:
\begin{itemize}
    \item \textbf{Agent:} The agent is a Deep Neural Network (DNN) comprising actor and critic networks which are responsible for  stabilizing learning and improve policy updates in time-varying network conditions.
    \item \textbf{State:} The state vector captures network indicators, including the aggregate buffer occupancy $\beta_o$, latency $\tau$, packet loss ratio $\rho$, and data rate $r$. To reduce dimensionality, averaged metrics are considered instead of per-user measurements.
    \item\textbf{Action:} The action represents the number RBs allocated for each slice. The total bandwidth is divided into $R$ RBGs leading to a discrete action space.
    \item\textbf{Reward:} The reward function evaluates the degree to which slice-level QoS requirements are satisfied. The agent aims to maximize the expected cumulative reward, thereby learning a resource slicing policy that maintains QoS compliance under dynamic wireless environments.
\end{itemize}

At each time slot $t$, the actor network observes the network state $s_t$ and selects a slicing action $a_t$ according to a stochastic policy $\pi(\cdot|s_t)$. 
The stochastic nature of the policy promotes exploration by sampling actions from a parameterized distribution. The objective of the policy is to maximize a trade-off between the expected cumulative reward and an entropy regularization term, which encourages exploration. The optimal policy $\pi^*$ is represented as
\begin{equation}
\pi^* = 
\arg\max_{\pi} 
\mathbb{E} \left[
\sum_{t} \gamma^t 
\left( 
R(s_t,a_t,s_{t+1}) 
- \lambda \log \pi(a_t|s_t) 
\right)
\right],
\end{equation}
where $\gamma \in (0,1)$ denotes the discount factor and $\lambda$ is the entropy coefficient controlling the exploration-exploitation balance. The soft $Q$-function represents the expected long-term return under the entropy-regularized objective and is defined as
\begin{equation}
\begin{split}
&Q(s_t,a_t) = 
\mathbb{E} \Big[
R(s_t,a_t,s_{t+1})\\ 
&+ \gamma \big(
\min_{i=1,2} Q_{\theta_i}(s_{t+1},a_{t+1})
- \lambda \log \pi(a_{t+1}|s_{t+1})
\big)
\Big],  
\end{split}
\end{equation}
where $Q_{\theta_1}$ and $Q_{\theta_2}$ correspond to the two critic networks used to mitigate overestimation bias.

The reward function is formulated to penalize violations of the service-level constraints of each user. For the eMBB slice, the users require high data rate with no stringent requirements on latency and packet loss. For the XR (IBS) slice, the users demand high data rate, low packet loss and tight scynchronization between the video and haptic feedback for a satisfactory user experience. We define $\bar{r}_k$ and $\bar{\rho}_k$ as the average data rate and packet loss ratio across users within the slice $k$. For the XR slice $s$, we define $\bar{\tau}_v$ and $\bar{\tau}_h$ as the average latency for the video and haptic traffic, respectively. 
\subsubsection*{Data Rate Component}
The data rate reward is designed to penalize insufficient throughput and is expressed as
\begin{equation}
R_k^{r} =
\begin{cases}
-\dfrac{r_k^{0} - \bar{r}_k}{r_k^{0}}, & \text{if } \bar{r}_k < r_k^{0}, \\
0, & \text{otherwise},
\end{cases}
\end{equation}
where $r_k^{0}$ denotes the minimum required data rate for slice $k$.
\subsubsection*{Packet Loss Component}
Similarly, the packet loss penalty is defined as
\begin{equation}
R_s^{\rho} =
\begin{cases}
-\dfrac{\bar{\rho}_s - \rho_s^{0}}{1-\rho_s^{0}}, & \text{if } \bar{\rho}_s > \rho_s^{0}, \\
0, & \text{otherwise},
\end{cases}
\end{equation}
where $\rho_s^{0}$ is the maximum tolerable packet loss.
\subsubsection*{Latency Component}
For the IBSs, the XR users needs to  maintain a tight synchronization between the video and haptic modalities. Hence, for the $s$ slice, we consider a synchronization threshold, which is based on the relative latency between video and haptic traffic. The reward component for the video traffic is given as

\begin{equation}
R_{s_v}^{\tau} =
\begin{cases}
-\dfrac{|\bar{\tau}_v - \bar{\tau}_h| - \tau_{\text{sync}}}{\tau_{\text{sync}}}, 
& \text{if } |\bar{\tau}_v - \bar{\tau}_h| > \tau_{\text{sync}}, \\
0, & \text{otherwise},
\end{cases}
\end{equation}
where $\tau_{\text{sync}}$ represents the maximum acceptable delay difference between the two modalities that ensures seamless user experience. Human-subject experiments suggest that users begin to perceive cross-modal desynchronization when the delay between haptic and video signals exceeds approximately 50 ms~\cite{sync2}. Moreover, for the haptic traffic, we formulate the reward contribution $R_{s_h}^{\tau}$ that considers the buffer latency as follows
\begin{equation}
R_{s_h}^{\tau} =
\begin{cases}
-\dfrac{\hat{\tau} - \bar{\tau}_0^h}{\tau_{\text{max}}-\bar{\tau}_0^h}, 
& \text{if }  \hat{\tau} > \bar{\tau}_0^h, \\
0, & \text{otherwise},
\end{cases}
\end{equation}
where $\hat{\tau}$ is the average buffer latency of the haptic traffic, $\tau_{\text{max}}$ is the maximum buffer latency of a packet before it is discarded and ${\tau}_0^h$ is the target buffer latency. Then, the latency contribution for the XR slice is $R_{s}^{\tau}=R_{s_v}^{\tau}+R_{s_h}^{\tau}$. Finally, the total reward function is calculated as the sum of all the components.
\begin{equation}
\mathcal{R} = R_s^{\rho} + R_s^{\tau}+\sum_{k \in \{s,c\}} 
R_k^{r}.
\end{equation}

Following the inter-slice allocation performed by the DRL agent, the intra-slice scheduler distributes the allocated RBGs among the slice users. The intra-slice scheduler assigns RBGs to user $u$ proportionally to its buffer occupancy $\beta_o^{u}$, thereby prioritizing users with higher traffic demand. The RBGs alloctated for user $u$ is computed as
\begin{IEEEeqnarray}{rCl}
N_{rbg}^{u}
&=&
\left\lfloor
\frac{\beta_o^{u}}{\beta_o}
\, N_{rbg}^{k}
\right\rfloor,
\label{eq:intraslice}
\end{IEEEeqnarray}
where $N_{rbg}^{k}$ denotes the number of RBGs allocated to slice $k$ and $\beta_o$ represents the total buffer occupancy of slice $k$. Compared to conventional resource distribution methods which follow round-robin approach,  this proportional allocation mechanism dynamically assigns more resources to users experiencing higher buffer backlog.

\section{Performance Evaluation}
\label{results}
\begin{table}[t]
\centering
\caption{Summary of system-level simulation parameters}
\label{tab:sys_sim_params}
\renewcommand{\arraystretch}{1.08}
\setlength{\tabcolsep}{6pt}
\begin{tabular}{|l|l|}
\hline
\textbf{Parameter} & \textbf{Value/Setting} \\
\hline
\multicolumn{2}{|c|}{\textbf{General}} \\
\hline
Deployment layout & $50\,\mathrm{m}\times 50\,\mathrm{m}\times 3\,\mathrm{m}$ \\
Number of devices & 5 eMBB and 10 - 25 IBS \\
Total bandiwidth & 100 MHz  \\
number of RBs & 272  \\
TTI & 14 OFDM symbols\\
Channel model & InH (Open Office) \cite{channel} \\
Macro BS power & 31 dBm \\
AP power & 10 dBm \\
$r_c^{0}$ & 45 Mbps \\
\hline
\multicolumn{2}{|c|}{\textbf{Hyperparameters of the SAC algorithm}} \\
\hline
Number of hidden layers & 4\\
$\gamma$ & 0.9\\
Learning rate & 0.0001\\
Update rate & 0.005\\
Replay buffer Size & 1$\times$ $10^6$\\
Optimizer& Adam\\
Entropy coefficient $\lambda$ &  Auto-tuned\\
Batch size & 1024\\
\hline
\multicolumn{2}{|c|}{\textbf{XR Traffic Model}} \\
\hline
Packet arrival rate for video & 90 packets/s \\
Packet arrival rate for haptic data & 1000 packets/s \\
$r_s^{0}$ & 60 Mbps \\
$\rho_s^{0}$ & $1\times10^{-5}$ \\
$\tau_{\text{sync}}$& 50 ms\\
\hline
\end{tabular}
\end{table}
The performance of the proposed method is evaluated using a system-level simulator following the 3GPP methodology~\cite{simulation}. The simulation implements a macro base station centered at a height of 3m in a 50m $\times$ 50m environment. The spatial distribution of the IBSs follows a Thomas Cluster Process (TCP) with five cluster centers and a standard deviation of $2$~m for the distance between offspring points and their corresponding cluster centers~\cite{TCP}. Each IBS is modeled as a cylindrical region with radius $0.25$~m and height $1.9$~m, where the cylinder center corresponds to the IBS location. A minimum separation of $4$~m between cluster centers and $0.5$~m between IBS centers is enforced to avoid excessive spatial overlap. The minimum time unit considered in the scheduling process is a transmission time interval (TTI), which is equivalent to the duration of 14 OFDM symbols and represents the time to allocate the RBGs to each slice. A summary of the simulation parameters are listed in Table~\ref{tab:sys_sim_params}

For the DRL algorithm, we assume that both the actor and critic networks share a similar architecture. Each network consists of 4 hidden layers with 400, 300, 200, and 100 neurons, respectively. The actor network output is connected to a softmax layer to produce a probability distribution over discrete actions, whereas the critic networks use linear output layers without additional nonlinear activation. The discount factor is set to $\gamma = 0.9$, and the learning rate is $10^{-4}$. The target network parameters are updated using a soft-update mechanism with rate $0.005$.  Training is performed over 10 episodes, each comprising 10,000 time steps with each step is 1 ms. The initial 200 frames are excluded to avoid the impact of transient traffic behavior.  After training, the policy is evaluated over 100 independent episodes of equal length, We use the work in~\cite{base} as a baseline. We implement the same reward function used in~\cite{base} while using the same observation and action spaces in our proposed method.
    \begin{figure}[t!] 
		\centering
		\includegraphics[width= 0.82\linewidth]{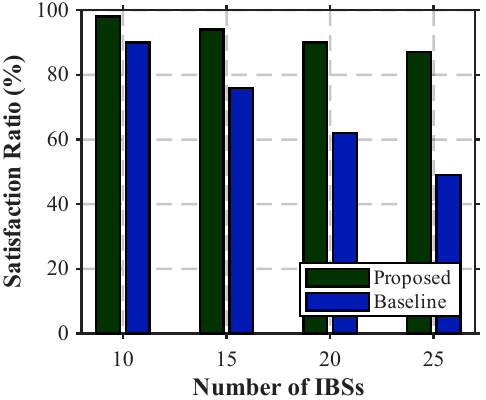} 
		\caption{Satisfaction ratio under varying number of IBS users.\label{SR}}
	\end{figure}

Fig.~\ref{SR} shows the satisfaction ratio under different number of IBSs. A user (XR or eMBB) is marked as satisfied if the corresponding QoS requirements are fulfilled over a single episode. Then, we calculate the satisfaction ratio using statistics aggregated from the results of 100 independent episodes. Based on the Wilson score interval for binomial proportions~\cite{wilson}, all the reported results have at least 95\% confidence interval. As depicted by Fig.~\ref{SR}, the satisfaction ratio of both methods gradually decreases as the number of IBSs increases due to intensified competition for radio resources and increased inter-slice contention. Nevertheless, our proposed slicing method consistently maintains a significantly higher satisfaction level across all density regimes. Specifically, at 10 IBS users, the proposed approach achieves approximately 98\% satisfaction, compared to around 85\% for the baseline. Under the most congested scenario with 25 IBS users, the proposed method still preserves about 87\% satisfaction, while the baseline falls sharply below 50\%. This corresponds to a relative improvement of approximately 7-74\% across the evaluated densities. The most significant degradation for the baseline was the violation of the synchronization latency between the video and haptic modalities, which has been efficiently handled by the proposed slicing method.
    \begin{figure}[t!] 
		\centering
		\includegraphics[width= 0.9\linewidth]{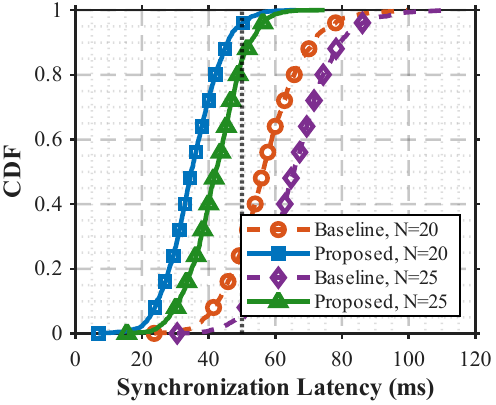} 
		\caption{CDF of the synchronization latency of the XR users with $N=20$ and $N=25$.\label{CDF}}
	\end{figure}
    \begin{figure}[t!] 
		\centering
		\includegraphics[width= 0.85\linewidth]{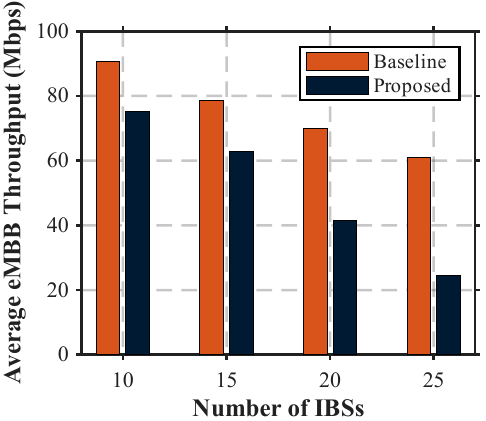} 
		\caption{Average throughput of eMBB users under varying number of IBS users.\label{eMBB}}
	\end{figure}

Fig.~\ref{CDF} depicts the cumulative distribution function (CDF) of the synchronization latency of the XR users. For the high IBS densities of 20 and 25, most of the values obtained from the proposed method fulfill the target synchronization latency while the baseline distribution shows a heavier tail toward higher latency values, exceeding the synchronization constraint of 50 ms. These results give more detailed insight on the effectiveness of the proposed method in maintaining acceptable user experience for XR users coexisting with other cellular users. 

In Fig.~\ref{eMBB}, we show the impact of the number of IBSs on the average throughput of the eMBB users. For the IBSs density of 10 and 15, both methods managed to maintain the average throughput above the target value of 45 Mbps. However, as the IBSs density increases to 20 and 25, the baseline exhibits noticeable degradation and fails to maintain the target throughput, while the proposed scheme still preserves a safe margin above 45 Mbps. Taking together, the obtained results demonstrate the effectiveness of the proposed SAC-based method in efficiently utilizing the available radio resources for supporting the emerging 6G IBSs, while maintaining acceptable QoS for the standard cellular users, 

\section{Conclusion}
\label{sec:conclusions}
The paper investigated a DRL-based method for RAN slicing, addressing the challenge of underlay coexistence scenario of IBSs and cellular users. We proposed a SAC-based slicing framework that dynamically allocates radio resources between XR and eMBB slices using a service-aware reward formulation. Performance evaluations have been conducted via system-level simulations and the obtained results showed that the proposed method achieves improved user experience compared to the baseline. Our findings therefore contribute to robust and scalable resource management solutions, supporting coexistence between heterogeneous services in future 6G network-of-networks architectures.
	\bibliographystyle{IEEEtran}

\bibliography{mybib}

\end{document}